\documentstyle[12pt]{article}
\title{Nonabelian Fields in Exact String Solutions}
\author {M. Z. Iofa and L. A. Pando Zayas\\
Nuclear Physics Institute \\
 Moscow State University\\
Moscow 119899, Russia }
\date{}
\def\intt{\int_{\Sigma}d^{2}z}
\def\pp{\partial_{z}}
\def\ppm{\partial_{\bar{z}}}
\def\pd{\partial_{\mu}}
\def\pu{\partial^{\mu}}
\def\T{1+r^2}
\def\Z{t^{2}-1}
\def\Am{A_{\bar{z}}}
\def\Ap{A_{z}}
\def\am{a_{\bar{z}}}
\def\ap{a_{z}}
\def\Qm{Q_{-}}
\def\Qp{Q_{+}}
\def\Fpm{F_{z\bar{z}}}
\def\P{\Phi}
\def\p{\phi}
\def\tg{\tilde{g}}
\def\tjp{\tilde{J}_{z}}
\def\tjm{\tilde{J}_{\bar{z}}}
\def\Jp{J_{z}}
\def\Im{I_{\bar{z}}}
\def\L{\Lambda}

\def\l{\lambda}

\def\te{\theta}

\begin{document}
\maketitle
\begin{abstract}
 Within the framework of "anomalously gauged"  Wess-Zumino-Witten (WZW) 
models, we construct solutions which include nonabelian fields. 
Both compact and noncompact groups are discussed.
In the case of compact groups, as an example of background containing 
nonabelian fields, we discuss conformal theory
on the $SO(4)/SO(3)$ coset, which is the natural generalization
of the 2D monopole theory corresponding to the $SO(3)/SO(2)$ coset.
In noncompact case, we  consider examples with $SO(2,1)/SO(1,1)$ and
$SO(3,2)/SO(3,1)$ cosets.
\end{abstract}

\section{Introduction}
 
The interest in exact string solutions for gravitational backgrounds, most of
which were obtained by using conformal field theories, has increased in
the last years (for a review see \cite{ts1} and references therein). 
The principal motivation for these studies was the hope that these
solutions could clarify some aspects
of black hole physics and would shed light on relation between the
world-sheet and the target-space dynamics. Gagued WZW models play a significant
role in  construction of exact solutions. Recently a class of theories
 based on
gauge extensions of the WZW models has been introduced \cite{hetco} in which
the gauge anomaly depends only on the gauge
field and is independent of a group element; furthermore, the anomaly has
the form of the two-dimensional chiral fermion anomaly \cite{witcom}. The
anomaly resulting from the gauged WZW model is then cancelled against the
one-loop anomaly of the fermions. The right sector of the model can be made
supersymmetric by an appropiate choice of  gauge couplings, left
fermions provide current algebra. In \cite{john1} it was shown that
 the monopole construction of paper \cite{gps} can be
 interpreted in terms of the anomaluosly gauged WZW model.

In this
paper the heterotic coset model technique is used to generate backgrounds
including
nonabelian fields. In sect. 2 we review the heterotic coset model approach
and
introduce a suitable parametrization for the $SO(3)/SO(2)$ coset. In sect. 3
we present a similar construction for the $SO(4)/SO(3)$ coset. The gauge
group $SO(3)$ being nonabelian, the resulting backgrounds are shown to 
include nonabelian gauge fields. In sect. 4 using parametrization of a group
element similar to that used in compact case, we consider examples of conformal
theories on $SO(2,1)/SO(1,1)$ and $SO(3,2)/SO(3,1)$ and cosets.
In sect. 5 we make some concluding remarks.

\section{Heterotic Coset Model Technique on the Example of 
$SO(3)/SO(2)$ Coset}

In this section we briefly describe construction of heterotic coset
models and discuss  potential problems that appear in the case of nonabelian
subgroups which are needed to introduce nonabelian fields as a part of
backgrounds. As is standard in the heterotic coset model construction, 
we begin with an
anomalously  gauged WZW model. In this paper we discuss
the right gauged WZW models, but further generalization is evident. The right 
gauged model is given  by the following action:

\begin{eqnarray}
I_{RGWZW}(g,A)&=&k I(g)+\frac{k}{2\pi}\intt Tr \Am g^{-1}\pp g
-\frac{k}{4\pi}\intt Tr \Ap\Am \nonumber \\
I(g)&=&\frac{1}{4\pi}\intt Tr \pd g\pu g^{-1} \nonumber \\
&+&\frac{1}{6\pi}\epsilon^{ijk}Tr\int\limits_{B,\partial B=\Sigma}
d^{3}y g^{-1}\partial_{i}gg^{-1}\partial_{j}gg^{-1}\partial_{k}g,
\end{eqnarray}
here $g\in G$,\quad  $A_{a}\in H \subset G$, \quad $k$ is the level of the 
Kac-Moody algebra.  Under the gauge transformations
\begin{equation}
g \to gh \quad \delta g=-gu \quad \delta A_a =-D_a u
=-\partial_a u-[A_a ,u],
\end{equation}
the action changes as
\begin{equation}
\delta I(g,A)=\frac{k}{4\pi} \int d^2 z Tr u(\pp\Am -\ppm\Ap)
\end{equation}

Since  this expression  has exactly the same structure as the nonabelian
chiral anomaly, we can add to the action $I(g,A)$ a set of fermions
and impose the condition of anomaly cancellation. Still we
have to keep in mind that since the anomaly (3) of the gauged WZW model is
classical, and the anomaly from the fermionic sector appears at the 
one-loop level, the anomaly
cancellation condition relates two different coupling constants. To
represent a resulting theory in a form of a
gauge invariant classical theory, we have to bosonize the fermionic sector.
As an example, let us consider conformal theory on the
$SO(3)/SO(2)$ coset. As a starting point for further generalizations, we
use  parametrization of the group element as in
\cite{bars}
\footnote
{In contrast to the commonly used Euler parametrization of the $SO(3)$
group, this parametrization can be can be generalized to the case of
higher dimensional groups.}. In the general case of $SO(d)/SO(d-1)$ cosets, the
group element of  $SO(d)$ can be represented as $g=th$
\footnote{To make the relation with the form of element $g$ of \cite{bars} 
explicit, we
have to identify $g$ with $g^{-1}$.}, 
where $h \in SO(d-1)$ and $ t \in
SO(d)/SO(d-1)$.  For the case of interest, the right gauged WZW,
a natural gauge choice is $h=1$ for which $g=t$.  The matrix $t$ is of the form:
$$
t=\left(
\begin{array}{cc}
b&(b+1)\hat{X^{\nu}}\\
-(b+1)\hat{X_{\mu}}& \delta_{\mu}^{\nu}-(b+1)\hat{X_{\mu}}\hat{X^{\nu}}
\end{array}
\right),
$$
here $\hat{X^{\mu}}$ is a row $(x^{1},x^{2})$, $\hat{X_{\mu}}$ is a column 
composed of the same elements,
$b=(1-x^{2})/(1+x^{2})$. The indices are
contracted with the Euclidean  metric $\delta_{\mu\nu}=diag(1,1)$.
Introducing two right-moving fermions which play the role of the
superpartners of the fields $x^{1}$ and $x^{2}$ and a pair of left-moving
fermions coupled to the gauge field with the strength $Q$, one obtains
the anomaly cancellation condition, which has the same form as in the
monopole theory \cite{john1},
$$
k=2(Q^{2}-1)=2Q_{+}Q_{-},
$$
here $Q_{\pm}=Q\pm 1$.  The coupling for the
right gauge field is fixed by requiring  supersymmetry in the right sector.
The gauge field in  equation (1) is  of the form
$$
\Ap =\ap T=\frac{1}{2}\ap\left(
\begin{array}{ccc}
0&0&0\\
0&0&1\\
0&-1&0
\end{array}
\right),
$$
where $T$ is the generator of the gauged subgroup $SO(2)\subset SO(3)$, and
the same relation holds for $\Am$ and $\am$.
The next step of this construction is bosonization of the fermionic
sector to obtain a classical gauge invariant action. The bosonic action is
required to yield the same classical "anomaly"  as the
(non)abelian one-loop fermionic anomaly and is taken in a form \cite{john1} 
\begin{equation}
I_{B}=\frac{1}{4\pi}\intt [(\pp \Phi -\Qp \ap)(\ppm \Phi -\Qp \am)-\Qm \Phi
\Fpm ].
\end{equation}
Under the transformation
$$
\delta A_{a}=\partial_{a}u \quad  \delta \P =Q_{+}u,
$$
the bosonic action  yields  classical anomaly of the form  
$$
\delta I_{B}=\frac{1}{4\pi}\intt Q_{+}Q_{-}F_{z\bar{z}}.
$$
which reproduces fermionic anomaly
\footnote{Another bosonic action which 
yields
the same anomaly is obtained by interchanging $Q_{+}$ and $Q_{-}$ in (4).
It can be verified that the final form of backgrounds is independent of
this ambiguity.}.

The same action (4) was used  in \cite{john1} for the treatment of the
monopole theory  on the $SU(2)/U(1)$ coset. At this point a remark is needed.
 Supersymmetry in the right
sector implies that the gauge field which gauges the
right symmetry in the WZW action must be used also for gauging  the right
fermions which are  the superpartners of the
group-manifold coordinates. For our case, we need two right fermions, the
number of left fermions is not constrainted, but if we  keep in mind that we
are going to bosonize
the fermions using a WZW model, we must take the same number of right and
left fermions, because one bosonic degree of freedom is represented by two
fermionic fields. Strictly speaking, the action (4) is not
the bosonized fermionic action, because the superpartners must belong to the
coset $SO(3)/SO(2)$, and we must use nonabelian bosonization
or, in other words, use a gauged WZW model defined on $SO(3)$. What
makes the above choice (4) admissible, is the assumption \cite{john1}
that the final metric must
depend only on coordinates of the bosonic part of the original
theory which we are going to
interpret as the coordinates of the physical space. The bosonic action  (4) is 
no longer suitable for nonabelian cosets, in which case the bosonized 
fermions must be
represented as an anomalously gauged WZW at level 1. In the next section we
will show that this fact
results in considerable complications in the problem of reading off the
metric.

Collecting (1) and (4), we get the total action
\begin{eqnarray}
I^{total}&=&kI(g)+\frac{k}{8\pi}\intt\frac{2}{k}\pp\P\ppm \P \nonumber \\
&+&\frac{k}{8\pi}\intt\left(\am(4\Jp-2\frac{(\Qp-\Qm)}{k}\pp\P)\right. \nonumber\\
&-&\left.\ap\frac{2(\Qp+\Qm)}{k}\ppm\P +\ap\am(2\frac{\Qp^{2}}{k}+1)\right).
\end{eqnarray}
Introducing new variables
\begin{equation}
x^{1}=r\cos\p \qquad x^{2}=r\sin\p,
\end{equation}
the term $I(g)$ and the current $J_{z}$ can be written as
\begin{eqnarray}
I(g)&=&\frac{2}{\pi}\intt\left(\frac{1}{(\T)^{2}}\pp r
\ppm r+\frac{r^{2}}{\T}\pp\p\ppm\p\right) \nonumber \\
\Jp&=&-\frac{2r^{2}}{\T}\pp\p.
\end{eqnarray}

Integrating out the gauge fields, we obtain the dilaton shift at the one-loop 
level,
which, in the present case, is trivial because it is independent of any field
variable and is equal to $\ln(2\Qp^{2}+k)$. For nonabelian
cosets and even for other gaugings, the behavior of the dilaton is completely
different.  The resulting action is:
\begin{equation}
I^{total}=kI(g)+\frac{k}{8\pi}\intt(\frac{1}{\Qp^{2}}\pp\P\ppm\P
+\frac{4}{\Qp}\Jp \ppm\P)
\end{equation}
To read off the metric, we still have to refermionize  this action. The naive
procedure for  reading off the metric of sigma models cannot be applied here
because the constants $k$ and $Q$ are connected by the anomaly cancellation
condition, and thus the fermionic sector affects the metric.  The method
proposed in \cite{john1} to extract the low-energy background consists in
rewriting the action (8) in a symmetric form which prepares
refermionization of the action.  This uniquely determines the change of the
metric of what we interpret as the bosonic sector of the heterotic sigma
model. Following this recipe, we have
\begin{eqnarray}
I^{total}&=&kI(g) \nonumber \\
&+&\frac{k}{8\pi}\intt\left(\frac{1}{\Qp^{2}}(\pp\P+2\Qp\Jp)
(\ppm\P+2\Qp\Im) \right. \nonumber \\
&-&\left.4\Jp\Im+\frac{2}{\Qp}(\Jp\ppm\P-\pp\P\Im) \right),
\end{eqnarray}
where $\Im=Tr Tg^{-1}\ppm g$. This form of the action makes the
way the metric is affected by fermions explicit.
Now it is possible to associate to  the total action the following heterotic
sigma model action
\begin{eqnarray}
I^{total}&=&\frac{k}{2\pi}\intt \left( G_{ij}\pd x^{i}\pu x^{j}
\right. \nonumber \\
&+&Tr \l_{R}(\ppm +A_{i}\ppm x^{i})\l_{R} \nonumber \\
&+&Tr \l_{L}(\pp +Q A_{i}\pp x^{i})\l_{L} \nonumber \\
&+& \left. F\l_{R}\l_{R}\l_{L}\l_{L} \right),
\end{eqnarray}
where the backgrounds are:
\begin{eqnarray}
G_{rr}&=&\frac{1}{(\T)^{2}}, \nonumber \\
G_{\p\p}&=&\frac{r^{2}}{\T}-\frac{r^{4}}{(\T)^{2}}, \nonumber \\
G_{\p r}&=&0,\nonumber \\
A_{\p}&=&2\frac{r^{2}}{\T},         \qquad A_{r}=0, \nonumber \\
F&=&\frac{1}{\Qp}.
\end{eqnarray}
To make contact with calculations of Johnson \cite{john1},
we must make the following change of coordinates:
$$
r=\tan\te,
$$
In these coordinates the part of the metric  from $I(g)$ is
$$
ds_{0}=d\te^{2}+\sin^{2}\te d\p^{2}.
$$
The modification of the metric resulting  from the fermions is
$$
-\frac{r^{4}}{(\T)^{2}}d\p^{2}=-\sin^{4}\te d\p^{2}
$$
Finally, rescaling  $\te\to\frac{\te}{2}$, we obtain  the  
background in the same form as in ref.  \cite{john1}:
\begin{eqnarray}
ds^{2}&=&d\te^{2}+\sin^{2}\te d\p^{2} \nonumber \\
A_{\p}&=&2\sin^{2}\frac{\te}{2}, \qquad A_{\te}=0.
\end{eqnarray}

In these coordinates, it is clear that the metric describes the two-dimensional
monopole theory. As we shall see in the next section, an advantage of the
parametrization we have used here is that the matrix form of the  group 
element is easily generalized for
higher dimensions. For higher-dimensional groups 
the form of the group element expressed in Euler angles is
very complicated, even for the group $SO(4)$ construction of the group element
in Euler parametrization is very involved.

\section{Nonabelian Fields from Coset SO(4)/SO(3)}

In this section  we apply the technique described in the previous section
to obtain backgrounds with  nonabelian fields.

As before, the group element is parametrized as $g=th$, where 
$X^{\mu}=(x^{1},x^{2},x^{3})$. As in section 2, we make transformation 
to spherical coordinates according to
\begin{eqnarray}
x^{1}&=&r\cos\te \nonumber \\
x^{2}&=&r\sin\te\cos\p \nonumber \\
x^{3}&=&r\sin\te\sin\p.
\end{eqnarray}
In these coordinates we have (a general property of this type of 
cosets is that the Wess-Zumino term is zero):
\begin{eqnarray}
I(g)&=&\frac{2}{\pi}\intt\frac{1}{\T}\left(\frac{1}{\T}\pd r \pu r \right.
\nonumber \\
&+&\left.r^{2}(\pd\te\pu\te+\sin^{2}\te\pd\p\pu\p\right).
\end{eqnarray}
The left-moving currents are defined as $J_{z}^a=Tr T^a g^{-1} \pp g$, 
where $T^a$ are the generators
of the gauge subgroup   $SO(3)$ used in this representation:
\begin{eqnarray}
T^{1}&=&\frac{1}{2}\left(
\begin{array}{cccc}
0&0&0&0 \\
0&0&1&0 \\
0&-1&0&0 \\
0&0&0&0
\end{array}
\right), \nonumber \\
T^{2}&=&\frac{1}{2}\left(
\begin{array}{cccc}
0&0&0&0 \\
0&0&0&1 \\
0&0&0&0 \\
0&-1&0&0
\end{array}
\right),    \nonumber \\
T^{3}&=&\frac{1}{2}\left(
\begin{array}{cccc}
0&0&0&0 \\
0&0&0&0\\
0&0&0&1 \\
0&0&-1&0
\end{array}
\right).
\end{eqnarray}

The currents are 
\begin{eqnarray}
J_{z}^1 & =& \frac{r^{2}}{\T}(-2\cos\p\pp\te +\sin\p\sin 2\te\pp\p), \nonumber \\
J_{z}^2 & =& \frac{r^{2}}{\T}(-2\sin\p\pp\te -\cos\p\sin 2\te\pp\p), \nonumber \\
J_{z}^3 & =& \frac{r^{2}}{\T}(-2\sin^{2}\te\pp\p).
\end{eqnarray}

Gauging of the model with respect to the subgroup $SO(3)$ requires three
gauge fields $A_{\mu}^{a}$.
The right fermions, which are the superpartnes of the fields $x_{i}$, are
minimally coupled to the gauge fields
and are  described by the action
\begin{equation}
I_F^R =\frac{k}{4\pi}\intt Tr \psi_{i} D_{\bar{z}}^{ij} \psi_{j},
\end{equation}
with $D_{\bar{z}}^{ij}=\ppm\delta^{ij}+ \Am^{a}(T^{a})^{ij}$.
This action produces the well-known chiral nonabelian anomaly which  in two
dimension is equal to \cite{zumino} (not to be confused with the abelian 
$\gamma^{D+1}$ anomaly):

\begin{equation}
\delta I_F^R =\frac{1}{4\pi}\intt Tr u (\pp\Am -\ppm\Ap).
 \end{equation}

The left fermions are 
minimally coupled to gauge fields with  arbitrary charges. Strictly
speaking, bosonizing this fermionic theory is a very nontrivial problem,
because, in principle, we must ensure conservation of all the
symmetries of the fermionic theory; moreover, we must guarantee that all the
correlation functions in both theories coincide.  Nevertheless, for the purpose 
of this paper,
we can use the arbitrariness  in construction of the heterotic coset model. 
If we postulate that the fundumental characteristic of a model is anomaly, 
then, since the expression for anomaly of the fermionic part of the original 
model is independent of fermionic fields,
we can introduce a bosonic theory yielding the same anomaly.  In the 
nonabelian case, we would like
to have something like the bosonic action (4), but this is no longer possible 
because the strength tensor now includes the
commutator of the gauge fields. Moreover, it may be checked  that there is
no modification of the action  (4) giving  the desired form of the anomaly.
A suitable possibilty for bosonization of the fermionic sector is to take
a gauged WZW action. The problem of this
choice is that for nonabelian subgroups it is impossible to express the
final background in a form in which backgrounds  depend only on what we would
like to call the target space geometry.  Actually, we end up with a mixture
of physical space coordinates and "internal" coordinates, because the
bosonic fields, that we introduced to replace the fermions, appear in 
backgrounds as internal coordinates.

To obtain  backgrounds, we must integrate out the gauge fields in 
the one-loop
approximation, i.e. in  the $k \to \infty$ limit (or in the $Q^{2} \to
\infty $ limit). In this limit, we can extract a part of the metric that
depends only on
"physical" coordinates which enter the original WZW model
\footnote{If we consider the action (5) as a 3D action on its own, 
independent of its
origin, then, since the 3D backgrounds are independent of the third
coordinate (the field $\Phi$), we can make the standard dimensional
reduction leading to 2D backgrounds which differ from those in (11). 
As noted in \cite{john1}, this way of doing is not correct because the 3D 
action was
obtained  by bosonizing the fermions and, as a result, the constants $k$ and
$Q$ are connected. Note, however, that backgrounds
obtained by these two ways have similar structure and differ only by
numerical coefficients.}.
 
In  the case under consideration, for bosonic action giving the
same gauge anomaly as the fermionic sector, we take the gauged WZW model
defined on the coset $SO(3)/SO(3)$ .  Our choice is dictated 
by the fact
that this coset has three bosonic fields (we need three right fermions) and
contains the gauge subgroup \footnote{ Since this theory has
anomalies,
it cannot be interpreted as a topological theory; in fact, the whole theory
is defined on the coset  $SO(4)\otimes SO(3)/SO(3)$.}. Here we see how the 
ambiguity in the choice of the bosonized fermionic theory can be used. 
The total action is:
\begin{eqnarray}
I^{total}&=&I_{RGWZW}+I(\tg)+\frac{1}{2\pi}\intt\left( (\am^{i}\tjp^{i}
-\ap^{i}Q^{i}\tjm^{i}) \right. \nonumber \\
&+&\left. \am^{i}\ap^{j}(Q^{j} Tr T^{j}\tg T^{i}
\tg^{-1}-\frac{1}{2}T^{i}T^{j}(1+Q^{i}Q^{j}) \right)
\end{eqnarray}
where there is no summation over the index of the charge $Q_{i}$ . Here
 $\tg \in SO(3)$ is represented by the $SO(4)$ matrix of the form 
$$
\left(
\begin{array}{cc}
1&0\\
0&\tg
\end{array}
\right)
$$
and the WZW action is defined as in (1).
Now we are going to specialize to a particular case. The
anomaly cancellation condition is
\begin{equation}
kTr T^{a}T^{b}+Tr T^{a}T^{b}-Q^{a}Q^{b}Tr T^{a}T^{b}=0
\end{equation}
A particular solution  is given by $Q^{1}=Q^{2}=Q^{3}=Q$. In this case, 
 the relation between $k$ and $Q$ is $k=\Qp \Qm$.
  Eliminating the gauge field in the semiclassical limit, we obtain
\begin{eqnarray}
I^{total}&=&kI(g)+I(\tg) \nonumber \\
&-& \frac{k}{2\pi}\intt(\frac{Q}{k^{2}}\tjm^{a}\L_{ab}\tjp^{b}
+\frac{Q}{k}\tjm^{a}\L_{ab}J_{z}^{a}) 
\end{eqnarray}
and 
\begin{equation}
\L^{ab}=Tr \left(\frac{1}{2k}T^{a}T^{b}(1+k+Q^{2})-
\frac{Q}{k}T^{b}\tg T^{a} \tg^{-1}
\right).
\end{equation}

Here the matrix $\L_{ab}$ is the inverse of  $\L^{ab}$. Nonabelian structure
of the action (21) is made explicit by writing the analog of the formula
(9), which  in this case is:
\begin{eqnarray}
I^{total}&=&kI(g)+I(\tg) \nonumber \\
&+&\frac{k}{2\pi}\intt\left(\frac{Q\Qm}{k^{2}}\tjm^{a}\L_{ab}\tjp^{b}\right.\nonumber \\
&-&\frac{Q^{2}}{k^{2}}(\tjm^{a}+\frac{k}{2Q}I_{\bar z}^{a})\L_{ab}
(\tjp^{b}+\frac{k}{2Q}J_{z}^{b}) \nonumber \\
&-&\frac{Q}{2k}(\tjm^{a}\L_{ab}J_{z}^{b}-I_{\bar z}^{a}\L_{ab}\tjp^{b})\nonumber \\
&+&\left. \frac{1}{4}I_{\bar z}^{a}\L_{ab}J_{z}^{b}\right),
\end{eqnarray}
where $I_{\bar z}^{a}=Tr T^{a}g^{-1}\ppm g$.
Comparing with the calculations of the previous section, we see that now the
role of $\ppm\P$ is played by $\tjm^{a}$. From equation (22) we 
have that in the large $k$ approximation
$$
\L^{ab}=Tr T^{a}T^{b}+O(\frac{1}{\sqrt{k}})
$$
In this limit, the part
depending on $\tg$ decouples from the part depending on $g$, leaving the
structure independent of the "inner" coodinates. 
\footnote{ In particular, this is the reason why we did not need the
explicit form of the currents $\tilde{J}^a$.}
As in section 2, we can
associate to (23) the heterotic model with the action 
\begin{eqnarray}
I^{total}&=&\frac{k}{2\pi}\intt \left( G_{ij}\pd x^{i} \pu x^{j}
\right. \nonumber \\
&+&Tr \l_{R}(\ppm \l_{R}+[A_{i}\ppm x^{i},\l_{R}]) \nonumber \\
&+&Tr \l_{L}(\pp \l_{L}+Q[A_{i}\pp x^{i},\l_{L}] \nonumber \\
&+& \left. F(g)\l_{R}\l_{R}\l_{L}\l_{L} \right).
\end{eqnarray}

Here we used the fact that the
right fermions are the superpartners of the group coordinates $x_{i}$. This
implies that the right gauge field may be introduced as $A^{a}_{i}:
Tr T^{a}g^{-1}\pp g=A^{a}_{i}\pp x^{i}$. Explicit expressions for the 
gauge fields are:
\begin{eqnarray}
A_{\p}^{1}&=&\frac{r^{2}}{\T}(\sin\p\sin 2\te), \nonumber \\
A_{\p}^{2}&=&\frac{r^{2}}{\T}(-\cos\p\sin 2\te), \nonumber \\
A_{\p}^{3}&=&\frac{r^{2}}{\T}(-2\sin^{2}\te), \nonumber \\
A_{\te}^{1}&=&\frac{r^{2}}{\T}(-2\cos\p), \nonumber \\
A_{\te}^{2}&=&\frac{r^{2}}{\T}(-2\sin\p), \nonumber \\
A_{\te}^{3}&=&0, \nonumber \\
A_{r}^{i}&=&0.
\end{eqnarray}
Here $i=1,2,3$. As before, the
antisymmetric tensor is zero. Using the same procedure as 
in the previous section, we obtain the following nonzero elements of the 
metric:
\begin{eqnarray}
G_{rr}&=& \frac{1}{(\T)^{2}} \nonumber \\
G_{\te\te}&=&\frac{r^{2}}{\T}-\frac{r^{4}}{(\T)^{2}} \nonumber \\
G_{\p\p}&=&\sin^{2}\te\left(\frac{r^{2}}{\T}-\frac{r^{4}}{(\T)^{2}}\right).
\end{eqnarray}

Changing the coordinate
$r\to\tan(r/2)$,  we obtain
\begin{eqnarray}
G_{rr}&=& 1 \nonumber \\
G_{\te\te}&=&\sin^{2}r \nonumber \\
G_{\p\p}&=&\sin^{2}r\sin^{2}\te.
\end{eqnarray}

This is the $S^{3}$ metric which emerges because of the relation between
the algebras $so(4)\sim so(3)\oplus so(3)$.
Models with this metric were discussed before
\cite{cosmo}. We can consider this theory as the spacial part of
a cosmological solution in a fashion similar to that of \cite{cosmo}, where
for the spacial part was taken the WZW model defined on the group $SO(3)$ 
and the
time coordinate was a free field with the corresponding conformal
field theory. 

\section{Noncompact Cosets}

In this section we apply  technique described in the preceding
sections to the noncompact cosets $SO(2,1)/SO(1,1)$ and $SO(3,2)/SO(3,1)$.
Here we omit some of the details discussed earlier. For the group 
$SO(2,1)$, fixing the gauge as in sect.2, we obtain the following form of 
element  

$$
t=\left(
\begin{array}{cc}
b&(b+1)\hat{X^{\nu}}\\
-(b+1)\hat{X_{\mu}}& \eta_{\mu}^{\nu}-(b+1)\hat{X_{\mu}}\hat{X^{\nu}}
\end{array}
\right),
$$
here $\hat{X^{\mu}}$ is a row $(x^{1},-x^{2})$, $\hat{X_{\mu}}$ is a column 
composed of the same elements, $b=(1-x^{2})/(1+x^{2})$. The indices are
contracted with the metric $\eta_{\mu\nu}=diag(1,-1)$. As usual, for the
noncompact cosets, we change the sign of the level $k$ in the gauged WZW
action. The anomaly cacellation condition remains the same, because the sign
of $Tr T^{2}$ also changes. The generator $T$ is 
$$
T=\frac{1}{2}\left(
\begin{array}{ccc}
0&0&0\\
0&0&1\\
0&1&0
\end{array}
\right).
$$
Changing the coordinates 
\begin{equation}
x^{1}=r\sinh t \qquad x^{2}=r\cosh t,
\end{equation}
and repeating the steps of described in section 2, we obtain the following
background
\begin{eqnarray}
ds^{2}&=&dr^{2}-\sinh^{2}rdt^{2} \nonumber \\
A_{t}&=&2\sinh^{2}\frac{r}{2} \qquad A_{r}=0.
\end{eqnarray}
This background can be interpreted as the electrically charged 2d black hole
(for  a more general construction see \cite{john1}).

The other coset of interest is $SO(3,2)/SO(3,1)$. Written in spherical
coordinates 
\begin{eqnarray}
x^{1}&=&t\sinh r\nonumber \\
x^{2}&=&t\cosh r\sin\te  \nonumber \\
x^{3}&=&t\cosh r\cos\te\sin\psi \nonumber \\
x^{4}&=&t\cosh r\cos\te\cos\psi,
\end{eqnarray}
the part of the metrifc calculated from $I(g)$ is 
\begin{equation}
ds^{2}_{0}=-\frac{8}{(\Z)^{2}}dt^{2}
+\frac{8t^{2}}{\Z}(dr^{2}+\sinh^{2}r(d\te^{2}+\sin^{2}\te d\psi^{2})).
\end{equation}
The generators of the gauged subgroup are $L_{ij}$ with $i,j=2,3,4,5$, where
$L_{ij}$ is the generator of the (psuedo)rotation in the plane $(i,j)$. The
gauge fields are calculated from the currents as in the previous sections;
here we present the final expresions for the currents 
\begin{eqnarray}
J_{z}^0 & =& \frac{t^{2}}{\Z}(-2\sin\te\pp r +\sinh 2r\cos\te\pp\te) \nonumber \\
J_{z}^1 & =& \frac{t^{2}}{\Z}(-2\cos\te\sin\psi\pp r-\sinh
2r\sin\te\sin\psi\pp\te+\sinh 2r\cos\te\cos\psi\pp\psi) \nonumber \\
J_{z}^2 & =& \frac{t^{2}}{\Z}(-2\cos\te\cos\psi\pp r-\sinh
2r\sin\te\cos\psi\pp\te-\sinh 2r\cos\te\sin\psi\pp\psi) \nonumber \\
J_{z}^3 & =& \frac{t^{2}}{\Z}(2\cosh^{2}r\sin\psi\pp\te+\cosh^{2}r\sin
2\te\cos\psi\pp\psi)\nonumber \\
J_{z}^4 & =& \frac{t^{2}}{\Z}(2\cosh^{2}r\cos\psi\pp\te+\cosh^{2}r\cos 
2\te\cos\psi\pp\psi)\nonumber \\
J_{z}^5 & =& \frac{t^{2}}{\Z}(2\cosh^{2}\cos^{2}\te\pp\psi)
\end{eqnarray}

As in the previous cases (symmetric spaces), the final metric takes the form
of the initial one after reparametrization of variables. The metric we
obtained is that of the anti-de Sitter space. This shows a substantial
difference from the treatment of the $SO(3,2)/SO(3,1)$ coset model in the
diagonal vector gauging of the WZW model \cite{cosmo}, where the low energy
limit of conformal field theory does not display any symmetry and cannot be
interpreted as a reasonable cosmological model.

\section{Conclusions}

We have constructed  backgrounds which were obtained by applying
heterotic coset model technique to cosets with  nonabelian subgroups.
As a further generalization of  this construction, we can consider
asymmetric gauging of both the left and right gauged subgroups. 

Since in the present approach string backgrounds are deduced from conformal
field theories, this ensures conformal symmetry of the corresponding
string theory.
More exactly, the required backgrounds are obtained by integrating out the
gauge fields. If this procedure could be carried out exactly,  one would 
obtain exact backgrounds valid in all orders in $\alpha'$.
 Otherwise, integration over the gauge fields is performed
in a certain approximation yielding backgrounds verifying the
$\beta$-equations in the corresponding order in $\alpha'$. Taking the 
limit $k\to\infty$ one
obtains backgrounds which are solutions of the leading-order
$\beta$-equations; however, in the limit $k\to\infty$ corresponding to
$\alpha'\to 0$, the effective action reduces to its leading part.

\section{Acknowledgements}

This investigation was partially supported by RFFI grant 96-02-16413.

\end{document}